\documentstyle[11pt,epsfig]{article}
\newcommand \be{\begin{equation}}
\newcommand \bea{\begin{eqnarray} \nonumber }
\newcommand \ee{\end{equation}}
\newcommand \eea{\end{eqnarray}}

\begin{document}

\title{\bf Increments of Uncorrelated Time Series Can Be Predicted With
a Universal $75\%$ Probability of Success}

\author{D. Sornette$^{1,2}$ and J.V. Andersen$^3$\\
$^1$ Laboratoire de Physique de la Mati\`{e}re Condens\'{e}e\\ CNRS UMR6622 and
Universit\'{e} de Nice-Sophia Antipolis\\ B.P. 71, Parc
Valrose, 06108 Nice Cedex 2, France\\
$^2$ Institute of Geophysics and
Planetary Physics and Department of Earth and Space Science\\ 
University of California, Los Angeles, California 90095\\
$^3$  Nordic Institute for Theoretical Physics\\
Blegdamsvej 17, DK-2100 Copenhagen, Denmark}

\date{\today}
\maketitle

\begin{abstract}
We present a simple and general result that the sign of the variations or increments of
uncorrelated times series are predictable with a remarkably high success probability of 
$75\%$ for symmetric sign distributions. The origin of this paradoxical result is
explained in details. We also present some tests on synthetic, financial and
global temperature time series.

\end{abstract}

\thispagestyle{empty}

\pagenumbering{arabic}

\section{Introduction}

Predicting the future evolution of a system from the analysis of
past time series is the quest of many disciplines, with
a wide range of useful potential applications including
natural hazards (volcanic eruptions, earthquakes, floods, 
hurricanes, global warming, etc.), medecine (epilectic seizure, cardiac arrest, 
parturition, etc.) and
stock markets (economic recessions, financial crashes, investments, etc.).
The absolute fundamental prerequisite is that the 
(possibly spatio-temporal) time series $x_1, x_2, ...$
possess some dependence of the future on the past.
If absent, the best prediction of the future is captured by the mathematical concept
of a martingale: the expectation ${\rm E}(x_{t+1}|{\rm past})$ of the future 
conditioned on the past is the last realisation $x_t$. In many applications, one is 
interested in the variation $x_{t+1}-x_t$ of the time series. 

The result we present below is, in one sense, obvious and, in another, quite counter-intuitive.
Starting from a completely uncorrelated time series, we know by definition that
future values cannot be better predicted than by random coin toss. However, we show
that the sign of the {\it increments} of future values can be predicted with a remarkably high
success rate of up to $75\%$ for symmetric time series. The derivation is straightforward
but the counter-intuitive result warrants, we believe, its exposition. This little 
exercice illustrates how tricky can be the assessment of predictive power and
statistical testing.

\section{First derivation}

Consider a time series $x(t)$ sampled at discrete
times $t_1, t_2, ...$ which can be equidistant or not. We denote $x_1, x_2, ...$ the
corresponding measurements. We assume that the measurements $x_1, x_2, ...$ are
i.i.d. (independent identically distributed).  Consider first the simple case where 
 $x_1, x_2, ...$ are uniformly and independently drawn in the interval $[0,1]$ and  
the average value or expectation is ${\rm E}(x) = 1/2$. 

\subsection{Prediction scheme}

We ask the following question:
based on previous values up to $x_i$, what is the best predictor for the increment
$x_{i+1}-x_i$?  A naive answer would be that, since the $x$'s are independent and
uncorrelated, their increments are also independent and the best predictor for 
the increment $x_{i+1}-x_i$ is zero (martingale choice). This turns out to be wrong.
If indeed the expectation of the increment is given by
\be
{\rm E}(x_{i+1}-x_i) = {\rm E}(x_{i+1}) - {\rm E}(x_{i}) = 1/2 - 1/2 =0~,
\ee
the conditional expectation ${\rm E}(x_{i+1}-x_i | x_i)$, conditionned on the
last realization $x_i$, is given by
\be
{\rm E}(x_{i+1}-x_i | x_i) = {\rm E}(x_{i+1} | x_i) - {\rm E}(x_i | x_i)
= {1 \over 2} - x_i~,   \label{akjalla}
\ee
where the term $1/2$ uses the independence between $x_{i+1}$ and $x_i$ 
(${\rm E}(x_{i+1} | x_i)= {\rm E}(x_{i+1})=1/2$) and the last
term in the r.h.s. uses the identity ${\rm E}(x_i | x_i) = x_i$.
We thus see that the sign of the increment has some predictability:
\begin{itemize}
\item if $x_i > 1/2$, the expectation is that $x_{i+1}$ will be smaller than $x_i$;
\item if $x_i < 1/2$, the expectation is that $x_{i+1}$ will be larger than $x_i$.
\end{itemize}
This predictability can be seen from the fact that the increments of $x(t)$ are
anti-correlated:
\be
{\rm E}\left(\left(x_{i+1}-x_i\right) \left(x_{i}-x_{i-1}\right)\right)
= {\rm E}(x_{i+1} x_i) - {\rm E}(x_{i+1} x_{i-1}) - {\rm E}(x_{i}^2) + {\rm E}(x_{i}x_{i-1})
={1 \over 4} - {1 \over 4} - {1 \over 2} + {1 \over 4} = - {1 \over 4}~.
\ee
This anti-correlation leads indeed to the predictability mentionned above, namely
that the best predictor for $x_{i+1}-x_i$ is that $x_{i+1}-x_i$ be of the sign opposite
to $x_{i}-1/2$.

Another way to understand where the predictability of the {\it increments} of incorrelated
variables comes from is to realize that increments are discrete realizations
of the differentiation operator. Under its action, a flat (white noise)
spectrum becomes colored towards
the ``blue'' (which is the opposite of the well-known action of integration which ``reddens''
white noise) and there is thus a short-range correlation appearing in the increments.

\subsection{Probability for a successful prediction}

\begin{figure}
\begin{center}
\epsfig{file=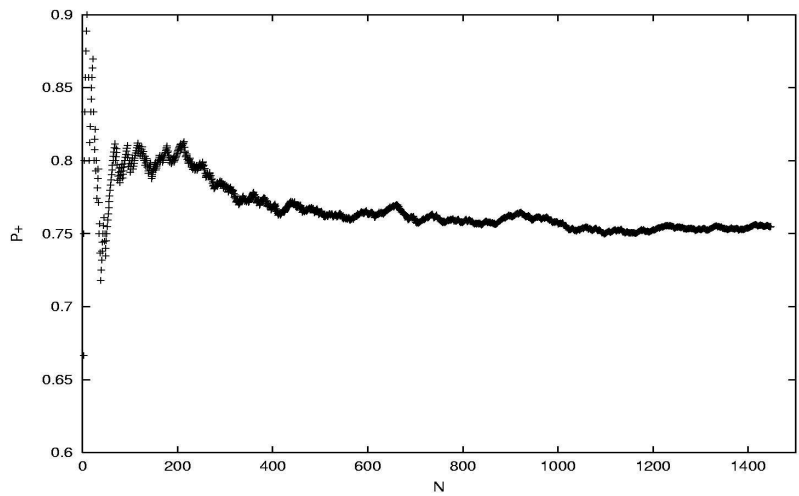,height=8cm,width=12cm}
\caption{\protect\label{fig1} Numerical simulation which evaluates $p_+$
as a function  of cumulative number of realisations
with the strategy that $x_{i+1}-x_i$ is predicted of the opposite sign 
to $x_{i}-1/2$,  using a pseudo-random number generator 
with values uniformely distributed between $0$ and $1$. }
\end{center}
\end{figure}

A natural question is to determine the success rate $p_+$ of this strategy, i.e. the 
probability that the sign of the increment $x_{i+1}-x_i$ be as predicted equal to the 
sign of $1/2 - x_i$. To address this question, we study the following quantity
\be
\epsilon \equiv
{\rm E}\left({\rm sign}\left(x_{i+1}-x_i\right) {\rm sign}\left({1 \over 2}-x_{i}\right)\right)~,
\label{fjajfjnn}
\ee
where the product of signs inside the expectation operator
is $+1$ if the prediction is born out by the data and $-1$ in the other case.
The relationship between $\epsilon$ and $p_+$ is
\be
\epsilon = (+1) p_+ + (-1) (1-p_+) = 2 p_+ -1 ~~~ \to ~~~p_+ = {1 \over 2} + {\epsilon \over 2}~.
\label{fjajka}
\ee
Expression (\ref{fjajka}) shows that $\epsilon$ quantifies the deviation for the
random coin toss result $p_+ = 50\%$. From the definition (\ref{fjajfjnn}), we have
\be
\epsilon = \int_0^{1 \over 2} dx_i ~(+1) \left( \int_0^{x_i} dx_{i+1} ~(-1) +
\int_{x_i}^1 dx_{i+1} ~(+1)\right) +
\int_{1 \over 2}^1 dx_i ~(-1) \left( \int_0^{x_i} dx_{i+1} ~(-1) +
\int_{x_i}^1 dx_{i+1} ~(+1)\right)~,
\ee
which gives 
\be
\epsilon = 1/2~~~~~{\rm and~thus}~~~~p_+ = 75\%~.   \label{dnanka}
\ee
Figure \ref{fig1} shows a numerical simulation which evaluates $p_+$ as a function 
of cumulative number of realisations
with the strategy that $x_{i+1}-x_i$ is predicted of the opposite sign 
to $x_{i}-1/2$,  using a pseudo-random number generator 
with values uniformely distributed between $0$ and $1$. 

\section{General derivation for arbitrary distributions}

This result is actually quite general. Consider an arbitrary random variable $x_i$
with arbitrary probability density distribution $P(x)$ with average
$\langle x \rangle$. We form the centered variable 
\be
\delta_i \equiv x_i - \langle x \rangle
\label{jfjjgfh}
\ee
with zero mean $\langle \delta \rangle = 0$
and pdf $P(\delta)$. Similarly to (\ref{akjalla}), we study the
conditional expectation of its increments $\delta_{i+1}-\delta_i$, given the
last realization $\delta_i$:
\be
{\rm E}(\delta_{i+1}-\delta_i | \delta_i) = {\rm E}(\delta_{i+1} | \delta_i) - 
{\rm E}(\delta_i | \delta_i) = - \delta_i~,   \label{akjalaaaala}
\ee
where we have used the fact that the $\delta_i$'s are uncorrelated. Thus, the best
predictor of the sign of the increment of the $\delta$'s is the opposite of the 
sign of the last realization.
We then quantify the probability of prediction success through the quantity
defined similarly to (\ref{fjajfjnn}) as
\be
\epsilon \equiv - {\rm E}\left({\rm sign}\left(\delta_{i+1}-\delta_i\right) 
{\rm sign}\left(\delta_{i}\right)\right)~,
\label{fjajfjnaan}
\ee
which is related to the success probability $p_+$ by (\ref{fjajka}).
It is easily calculated as
\bea
\epsilon &=& \int_{-\infty}^0 d\delta_i~P(\delta_i)~(+1)\left[ 
\int_{-\infty}^{\delta_i} d\delta_{i+1}~P(\delta_{i+1})~(-1) + 
\int_{\delta_i}^{+\infty} d\delta_{i+1}~P(\delta_{i+1})~(+1)\right] \\
&+& \int_0^{+\infty} d\delta_i~P(\delta_i)~(-1)\left[ 
\int_{-\infty}^{\delta_i} d\delta_{i+1}~P(\delta_{i+1})~(-1) + 
\int_{\delta_i}^{+\infty} d\delta_{i+1}~P(\delta_{i+1})~(+1)\right] ~.  \label{hfhhfk}
\eea
It is convenient to introduce the cumulative distribution
\be
F(x) \equiv \int_{-\infty}^{x} d\delta ~P(\delta)~,
\ee
and the probabilities $F_- = F(0)$ (resp. $F_+ = 1-F(0)$) that $\delta$ be less
(resp. larger) than $0$. Expression (\ref{hfhhfk}) transforms into
\be
\epsilon = F_- - F_+ - \left( F(0) \right)^2 +  \left( F(+\infty) \right)^2 -  \left( F(0) \right)^2~,
\ee
where we have used the identity $ 2\int_{-\infty}^y dx~P(x)~F(x) = [F(y)]^2$. Using
the definition of $F_-$ and $F_+$ and the normalization $F(+\infty)=1$ leads to
\be
\epsilon = 2 F_+(1-F_+)~~~~~{\rm and}~~~~~p_+= {1 \over 2}+ F_+(1-F_+)~.  \label{jfhhhhjs}
\ee
For symmetric distributions and for those distributions such that $F_+=1/2$, we
retrieve the previous result (\ref{dnanka}). This result
is thus seen to be very general and independent of the shape of the distribution of the
i.i.d. variables as long as $F_+=1/2$ (attained in particular but not exclusively
for symmetric distributions). Note that the value $p_+=75\%$ is
the largest possible result attained for $F_+=1/2$. For $F_+\neq1/2$, $0.5 \leq p_+ < 0.75$.

\begin{figure}
\begin{center}
\epsfig{file=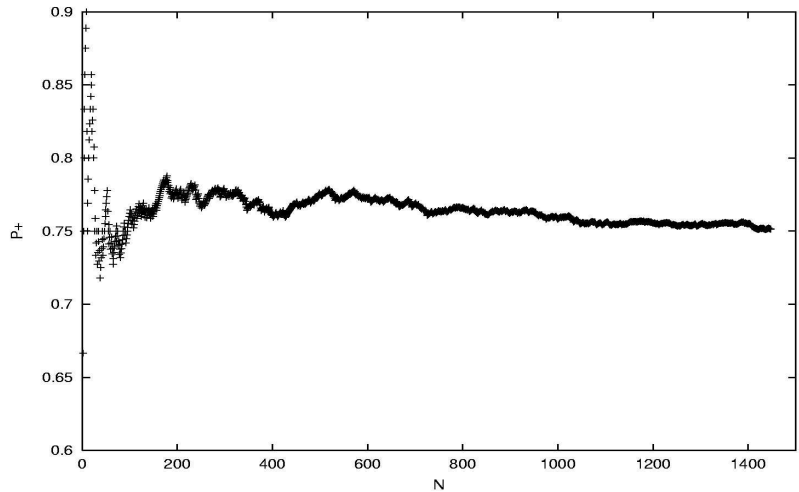,height=8cm,width=12cm}
\caption{\protect\label{fig2} Predictability of the increments of the price
variations, i.e. the acceleration, of
the thirty year US treasury bond TYX from Oct. 29, 1993 till  Aug. 9,
1999. Specifically, we start from the daily close quotes $q(t)$ and construct 
the price variations $\delta q(t) = q(t)-q(t-1)$. We try to predict the variation of
 $\delta q(t)$ with 
the strategy that $\delta q(t+1)-\delta q(t)$ is predicted of the opposite sign 
to $\delta q(t)-\langle q \rangle$. The corresponding success probability $p_+$ 
is plotted as a function of
time by cumulating the realizations to estimate $p_+$.}
\end{center}
\end{figure}

Figure \ref{fig2} shows the estimation of $p_+$
  used on the thirty year US treasury bond TYX from Oct. 29, 1993 till  Aug. 9,
1999. Specifically, we start from the daily close quotes $q(t)$ and construct 
the price variations $\delta q(t) = q(t)-q(t-1)$. We try to predict the variation of
 $\delta q(t)$ with 
the strategy that $\delta q(t+1)-\delta q(t)$ is predicted of the opposite sign 
to $\delta q(t)- \langle \delta q \rangle$. 
The corresponding success probability $p_+$ is plotted as a function of
time by cumulating the realizations to estimate $p_+$. As expected, at the beginning, 
large fluctuations express the limited statistics. As the statistics improves, $p_+$ 
converges to the predicted value $75\%$. We note that, in comparison to the 
pseudo-random number series shown in figure \ref{fig1}, the convergence seems to occur
at a similar rate, suggesting that there are no appreciable global short-range
correlations, in agreement with many previous statistical tests \cite{Loetal,dsBP,MantegnaStanley}.

\section{Discussion}

This paradoxical result tells us
that one can get on average a success rate of three out of four
in guessing what is the sign of the increment of uncorrelated random variables. This is
quite surprising a priori but, as we explained above, stems from the action of the 
differential operator which makes the spectrum ``blueish'', thus introducing short-range
correlations. 

This predictive skill does not lead to any anomalies. Consider for instance the 
time series of price returns of a stock market. According to the efficient market hypothesis
(ref.\cite{Loetal} and references therein)
and the random walk model, successive (say daily) price returns of liquid organized markets are 
essentially
independent with approximately symmetric distributions. Our result (\ref{jfhhhhjs})
shows then that we can predict with a $75\%$ accuracy the sign of the {\it increment}
of the daily returns (and not the sign of the returns that are proportional to the 
increment of the prices themselves). This predictive skill is {\it not} associate to an
arbitrage opportunities in market trading. This can be seen as follows.
For simplicity of language, we consider 
price returns $\delta$'s relative to their average so that we deal with uncorrelated variables
with zero mean as defined in (\ref{jfjjgfh}). In addition, we restrict our discussion
to the optimal case where $F_+=1/2$. Consider first the situation where
$\delta_i$ is positive and quite large (say two standard deviations above zero).  
We expect that any typical realization, and in particular the next one
$\delta_{i+1}$, to be positive or negative but close to zero to within say one standard
deviation. This implies that we expect with a large probability
$\delta_{i+1}$ to be smaller than $\delta_{i}$. This is the guess that
is compatible and in fact constructs the result (\ref{jfhhhhjs}). Consider now
the second situation where $\delta_i$ is positive but very small and close to zero.
We then have by construction of the process that $\delta_{i+1}$ will be larger or smaller
than $\delta_i$ with probability close to $1/2$. In this case, we loose any predictive skill.
What the result (\ref{jfhhhhjs}) quantifies
mathematically is that all these types of realizations averages out to a global
probability of $75\%$ for the sign of the increment to be predicted by the sign of $-\delta_i$.
This large value is not giving us any ``martingale'' (in the common sense of the word). 
Actually, it
states simply that, for independent realizations, 
large values have to be followed by smaller ones. 
This analysis relies fundamentally on the independence between successive
occurrence of the variables $\delta_i$. Predicting with $75\%$ probability the sign
of $\delta_{i+1}-\delta_i$
does not improve in any way our success rate for prediction the sign of $\delta_{i+1}$
(which would be the real arbitrage opportunity).

\begin{figure}
\begin{center}
\epsfig{file=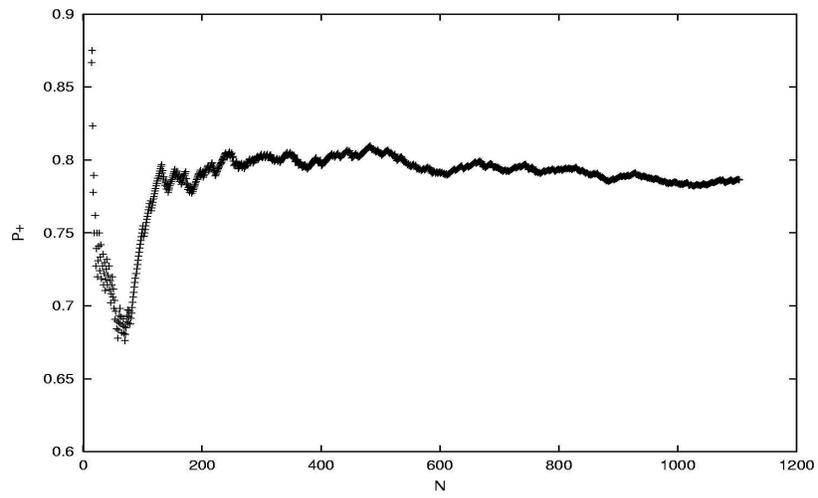,height=8cm,width=12cm}
\caption{\protect\label{fig3}  Success probability $p_+$ for the prediction of the variations
of the isotopic deuterium time series from the Vostok (south Pole)
ice core sample,
which is a proxy for the local temperature from about 220 thousand years in the past to present.
The data is taken from \cite{datadeut}. }
\end{center}
\end{figure}

Deviations from $p_+ = 75\%$, and in particular results larger than $75\%$ which 
is a maximum in the uncorrelated case (see (\ref{jfhhhhjs}), 
signal the presence of correlations. An instance 
is shown in figure \ref{fig3} which plots $p_+$ for the prediction of the variations
of the isotopic deuterium time series from the Vostok (south Pole) ice core sample,
which is a proxy for the local temperature from about 220 ky in the past to present.
The data is taken from \cite{datadeut}. We observe that $p_+$ remains 
above $75\%$ showing a significant genuine anti-correlation.

Acknowledgements: We thank P. Yiou for providing the temperature time series and S. Gluzman
for stimulating discussions.

\pagebreak

 \end{document}